**Honeycomb-structure RuI$_3$, a new quantum material related to *α*-RuCl$_3$**

*Danrui Ni, Xin Gui, Kelly M. Powderly, and R. J. Cava\**
Department of Chemistry, Princeton University, Princeton, NJ 08544, USA
E-mail: rcava@princeton.edu

**Abstract**

The Kitaev model predicts that quantum spin liquids (QSLs) will form at low temperatures under certain special conditions, and materials hosting the QSL state are frequently sought. The layered honeycomb lattice material *α*-RuCl$_3$ has emerged as a prime candidate for displaying the Kitaev QSL state. Here we describe a new polymorph of RuI$_3$ with a layered honeycomb lattice structure, synthesized at moderately high pressures and stable under ambient conditions. Preliminary characterization reveals metallic, paramagnetic behavior, the absence of long-range magnetic order down to 0.35 K and an unusually large *T*-linear contribution to the heat capacity at low temperatures. We propose that this RuI$_3$ phase, with a layered honeycomb lattice and strong spin-orbit coupling, provides a new route for the characterization of quantum materials.

## 1. Introduction

The Quantum spin liquid (QSL) state is a highly uncommon topological state of matter in which long-range entangled spins remain entangled even at absolute zero temperature.[1–3] This gives rise to an extensive residual spin entropy due to many-body entanglement, and allows for the possibility of some remarkable properties, such as the preservation of quantum information against decoherence.[4–6] The absence of conventional long-range magnetic ordering in QSLs does not necessarily rely on geometric frustration of the spin ordering - instead it can be derived from quantum fluctuations of spins on lattices that are not usually considered frustrated.[6,7] QSLs can arise in some latices due to strong bond-dependent spin anisotropy, yielding frustrated spin configurations on a single site, as postulated in the Kitaev model.[8,9] The layered honeycomb lattice is an excellent potential host for this state, leading to the so-called Kitaev QSL, which has recently attracted considerable attention both theoretically and experimentally. Materials with $S = 1/2$ spins arrayed on a honeycomb lattice and significant spin-orbit-coupling (SOC)[2,10] are particularly good candidates for hosting this state.

$α$-RuCl$_3$, as a promising KQSL candidate based on a $4d$ transition metal-based honeycomb lattice, has been the subject of a great number of investigations. Its unconventional behavior, including a highly unusual magnetic excitation spectrum as well as the emergence of Majorana fermions, has been revealed by magnetization, heat capacity, neutron scattering and thermal transport measurements.[3,11–13] Its ground state seems to be close to the ideal Kitaev QSL state,[13–15] in spite of the magnetic ordering of $α$-RuCl$_3$ in zero applied magnetic field and the stacking faults that are sometimes reported in its crystal structure. In the meantime, the iodine-based version of RuCl$_3$, honeycomb structured RuI$_3$ which until now has not been known to exist, has been the subject of several theoretical studies[16,17] of its monolayer form. The compound

RuI$_3$ has been reported, but studies have shown it to crystallize in a 1D chain structure instead of in honeycomb layers.[18,19] Here, we report a high-pressure solid-state synthesis of RuI$_3$ in a honeycomb-layered structure ($\alpha$-RuI$_3$). Preliminary characterization is conducted on the physical properties of the compound, which show it to be a paramagnetic metal with a relatively high T-linear contribution to its heat capacity, and density functional theory calculations are carried out on the electronic band structure in order to learn about the electronic factors that may influence its electronic and magnetic state.

## 2. Results and Discussion

To synthesize the layered honeycomb structure of RuI$_3$, commercially available amorphous RuI$_3$ powder (Alfa Aesar), was heated under a pressure of 6 GPa at 800 °C for one hour. The resulting honeycomb structure phase of RuI$_3$ is stable at ambient pressure for at least a few days in air before hydrating - appearing to be less hygroscopic than honeycomb structure RuCl$_3$ powder. The vapor transport method was employed on post-synthesis samples in a quartz tube sealed under vacuum to purify the product, with the hot end at 300 to 350 °C and the cold end at ambient temperature. A small amount of chemical impurity, presumably present due to its presence in the starting material, were transferred to the cold end, and honeycomb structure $\alpha$-RuI$_3$ is maintained in the hot zone with no signs of decomposition or phase transformation. This suggests that layered-honeycomb-structure $\alpha$-RuI$_3$ is stable for at least a few hours at 350 °C under static vacuum.

The crystal structure of the material was determined by single crystal X-ray diffraction (SXRD) at ambient temperature, with single crystals picked from the post-reaction samples. The material has a three-layer centrosymmetric rhombohedral symmetry structure, with a honeycomb-layer structure in space group $R\bar{3}$ (No. 148, Figure 1). At 300 K the cell parameters are $a = 6.7913(6)$

Å, and $c$ = 19.026(3) Å. The atomic coordinates are listed in Table 1, and further crystallographic data are presented in Table S1 to S3. A structural detail is that the normally empty interstitial sites (Ru2, Wyckoff position $3a$) in the honeycomb lattice appear to be occupied by a very small number of Ru atoms. As diffraction measurements are a positional average over the whole crystal, the very small partial occupancy of Ru at position $3a$ is likely due to the presence of a small number of stacking faults,[20] such has often been observed in RuCl3. Without constraints among the occupancy parameters in the structure refinement, the honeycomb material freely refines to the slightly Ru deficient formula $Ru_{0.97}I_3$; the ideal $6c$ site of the honeycomb lattice is about 96% occupied while the ideally vacant $3a$ site is about 2% occupied. Detailed structural study, especially by high resolution electron microscopy, which can observe stacking faults or other structural errors, if present, may be of future interest.

Each $RuI_6$ octahedron in the $\alpha$-$RuI_3$ structure is close to an ideal geometry, with Ru-I bond lengths in a narrow range between 2.67 and 2.68 Å, and I-Ru-I bond angles of 86° to 92°. This suggests that distortions from octahedral crystal fields are likely to be small. As shown in Figure 1C, each unit cell contains three honeycomb layers in the usual A B C rhombohedral geometry. The honeycomb layers are built of edge-sharing $RuI_6$ octahedra at a Ru-Ru distance of 3.9 Å. The distance between the layers is approximately 6.3 Å. This interlayer separation is larger than the ones in $P3_112$ symmetry[21] $\alpha$-$RuCl_3$ (5.7 Å), suggesting that there will likely be weaker magnetic coupling between layers in $\alpha$-$RuI_3$ than is found in $\alpha$-$RuCl_3$. No structural phase transition was observed for honeycomb-structure $RuI_3$ single crystals between 100 and 300 K.

In Figure 2, powder XRD confirms the consistency of the bulk product with the honeycomb structure observed by single crystal diffraction. As shown in the inset, the layered nature of honeycomb lattice $\alpha$-$RuI_3$ can be straightforwardly seen by visual inspection of a scanning electron microscope (SEM) image, while energy-dispersive X-ray (EDX) measurements verify the 1 to 3 Ru to I ratio and the uniform element distribution seen in the single crystal XRD study. Our material is quite different from that previously reported for RuI$_3$, which was said to adopt a 1D chain structure ($\beta$-RuI$_3$).[18,19,22] The successful synthesis by a modest high pressure method of a new layered honeycomb phase of RuI$_3$, which has not been previously experimentally reported, may be because the layered honeycomb lattice tends to be preferred under pressure, as the relative molar volumes of 1D chain $\beta$-RuI$_3$ and honeycomb structure $\alpha$-RuI$_3$ are 131.5 Å$^3$ and 126.7 Å$^3$ respectively, suggesting that pressure drives the system towards a slightly more condensed phase (around 4% higher in density). Alternatively, it may be that the high pressure apparatus simply prevents the escape of iodine from the system and higher pressure is not actually a factor in the compound's thermodynamic stability. That being said, the amorphous RuI$_3$ starting material starts to release I$_2$ on heating above 200 °C under a static vacuum in a quartz glass tube, which may preclude a possible ambient pressure phase transition as from $\beta$-RuCl$_3$ to $\alpha$-RuCl$_3$ on increasing the temperature,[3] and thus a system that can prevent iodine from evolving away at lower or ambient pressure may also result in the preparation of honeycomb structured $\alpha$-RuI$_3$. This newly discovered RuI$_3$ phase will be of great interest due to its heavy-metal based honeycomb lattice, which allows for the significant chemical factors such as the Ru-Cl vs. Ru-I orbital hybridization, to the appearance of potentially unusual physics to be tested. The elementary physical properties of $\alpha$-RuI$_3$ were characterized using a Quantum Design Dynacool Physical Property Measurement System (PPMS). Magnetic susceptibility (defined as *M*/*H*) was derived from the measured magnetization (*M*) on powder samples at temperatures between 1.8 and 300 K under a 1 kOe applied magnetic field (*H*). The result is shown in Figure 3A. Honeycomb structured $\alpha$-RuI$_3$ shows relatively weak paramagnetic behavior over a wide temperature range and a low temperature upturn, with no three-dimensional magnetic ordering features visible down to 1.8 K. The magnetization shows a linear dependence on the external field from -9 T to 9 T at 200 K (Figure 3A inset) and an S-shaped character at 2 K. Heat capacity ($C_p$) was measured on a powder-pressed dense pellet of honeycomb structured RuI$_3$ between 10 and 0.35 K and no phase transition was observed. $C_p / T$ is plotted versus $T$ in the main panel of Figure 3B. The upward tail visible below 1 K is likely to be dominated by the nuclear quadrupole contribution from Ru, which has been observed and reported for other Ru containing compounds such as RuO$_2$,[23] though that contribution may mask the presence of a very low temperature magnetic transition. The total heat capacity of a quantum material ($C_{total}$) at low temperature can be described as the sum of electronic, magnetic and phononic contributions, e.g. $C_{total} = C_{electron} + C_{mag} + C_{phonon}$, where electronic or other types of linear-$T$ contributions can be of the form $\gamma T$, in addition to those of the phonons: $C_{phonon} = \beta_1 T^3 + \beta_2 T^5$.[24] Thus by fitting the temperature range above any potential magnetic transition using a polynomial equation (shown in the main panel of Figure 3B), the values obtained are 0.0293, 0.0472, and -1.2 x 10$^{-5}$ J/mol-K respectively for $\gamma$, $\beta_1$, and $\beta_2$. This result is distinctive compared to that seen for most QSL candidate materials as a $\gamma T$ term is rarely observed in heat capacity fitting,[3] and it is also obviously larger than the common $T$ linear terms of metal trihalides (for example, for layered material CrI$_3$,[25] the linear-$T$ term is reported to be 1.17 mJ mol$^{-1}$ K$^{-2}$). The relatively large magnitude of the gamma term, $\gamma$ = 29.3 mJ mol$^{-1}$ K$^{-2}$, is surprising and suggests that some type of unconventional behavior may be going on in the electronic or spin system of this material at low temperature.

The low temperature upturn in the magnetic susceptibility shown in Figure 3A is present in all our preparations of this material and seems to be too large to be due to the presence of spins that are not coupled to the remainder of the system (estimated from the measured magnetic susceptibility, via a Curie Weiss Law fitted from 1.8 to 10 K, to be about 30% of a spin 1/2 contribution, or $\mu_{eff}$ = 0.53 per formula unit, with $\theta$ = -3 K). Thus, the observation of no long-range magnetic ordering in $\alpha$-RuI$_3$ down to 0.35 K means either that the spin ordering is frustrated or that there is some type of 2D magnetic ordering present that is not visible as a feature in the susceptibility or heat capacity. This is in contrast to the case in $\alpha$-RuCl$_3$, where a long-range magnetic ordering transition with a Néel temperature $T_N$ = 7 K is observed for bulk materials.[11,26,27] The Ru$^{3+}$ ions on which both RuCl$_3$ and RuI$_3$ are based, with $4d^5$ configuration, are in a $j_{eff}$ = 1/2 state due to strong SOC,[3] and when compared with their larger moment counterparts, $j$ = 1/2 moments are more amenable to the influence of quantum fluctuations.[3] Compared with $\alpha$-RuCl$_3$, which develops long-range antiferromagnetic order, the newly discovered $\alpha$-RuI$_3$, with no long-range order down to 0.35 K in zero applied field, may require either 0 applied magnetic field or much weaker applied magnetic field in order to access the potential QSL behavior reported[12,15] for $\alpha$-RuCl$_3$.

A preliminary resistivity measurement was conducted on an as-made dense piece of RuI$_3$ directly from the high-pressure synthesis furnace, and metallic behavior was observed between 1.8 and 240 K (Figure 3B inset). This again is in distinct contrast to what is observed in RuCl$_3$, where semiconducting behavior is seen.[26–28] A weak feature observed between about 260-280 K in the resistivity is not shown in the figure as in measurements on multiple samples it appears to depend on the presence of adsorbed water on the sample surface.

Finally, DFT-based calculations on the electronic structure of $\alpha$-RuI$_3$ were performed with and without SOC and the Coulomb repulsion U terms (U = 4 eV, employed for Ru $d$ electrons based on the value used in such calculations for $\alpha$-RuCl$_3$.[29] Lower values of U also lead to a calculated metallic electronic structure.) The results are shown in Figure 4 with comparison to the calculated electronic band structures of $P3_112$ $\alpha$-RuCl$_3$. Both SOC and non-SOC-calculated band structures of $\alpha$-RuI$_3$ suggest that the compound should be metallic - the band structure shows obvious changes with both SOC and U terms included, but the metallic state near the Fermi Energy is maintained, consistent with the preliminary resistivity measurement. When both SOC and U are included in the calculation, $\alpha$-RuCl$_3$ shows a semiconducting band structure, consistent with experimental observations. The semiconducting nature of $\alpha$-RuCl$_3$ that we find in the calculation is also consistent with previous reports.[30]

## 3. Conclusions

A new phase of RuI$_3$, displaying a layered honeycomb structure, is reported. Synthesized at moderate pressures, it is stable under ambient conditions in lab air. Preliminary characterization indicates that it differs from RuCl$_3$ in three primary ways: (1) it does not show any signs of long-range magnetic ordering down to 1.8 K under applied fields of 1000 Oe or to 0.35 K with zero field, (2) it has a surprisingly large linear-$T$ contribution to the heat capacity compared to other QSL candidates, and (3) it is a metallic conductor. The spin-1/2 honeycomb lattice as well as the significant SOC of $\alpha$-RuI$_3$ make it a viable candidate for further study as a Kitaev QSL candidate. We argue that $\alpha$-RuI$_3$ can serve as a new platform for exploring the consequences of the Kitaev spin model, especially when its properties are compared to those of RuCl$_3$.

**Experimental Section**

Amorphous RuI$_3$ starting material was purchased from Alfa Aesar (anhydrous, ≥ 99%). The powder was loaded in a boron nitride crucible which was then inserted into a pyrophillite cube assembly. The system was pressed to 6 GPa using a cubic multi-anvil system (Rockland Research Corporation) and heated to 800 °C at 50 °C/min with temperature determined by an internal thermocouple. The sample was kept at 800 °C for 1 hour and then quench-cooled before decompression. The product obtained displayed a black color and was relatively stable in air. Small single crystals could be isolated in the post-reaction samples and were used in the single crystal Xray diffraction (SXRD) characterization of the structure.

Multiple $RuI_3$ crystals (~90×90×20 µm$^3$) were studied by SXRD to determine the crystal structure of the new material. The structure, consistent among all crystals, was determined using a Bruker D8 QUEST diffractometer equipped with APEX III software and Mo radiation ($\lambda$= 0.71073 Å) at room temperature. The crystals were mounted on a Kapton loop. Data acquisition was made *via* the Bruker SMART software with corrections for Lorentz and polarization effects included. A numerical absorption correction based on crystal-face-indexing was applied through the use of *XPREP*. The direct method and full-matrix least-squares on F$^2$ procedure within the SHELXTL package were employed to solve the crystal structure.[31,32] PXRD patterns were collected using a Bruker D8 Advance Eco with Cu Kα radiation ($\lambda$= 1.5406 Å). The Le Bail fitting of the acquired XRD patterns was conducted via the TOPAS software.

Magnetization and heat capacity measurements were carried out using a Quantum Design PPMS (Dynacool), equipped with a vibrating sample magnetometer (VSM) option. Resistivity data were also collected on QD PPMS Dynacool, with platinum wires attached to samples using DuPont 4922N silver paint.

The electronic structure and electronic density of states (DOS) of $RuI_3$ were calculated using the WIEN2k program.[33,34] The electron exchange-correlation potential used to treat the electron correlation was the generalized gradient approximation.[35] The conjugate gradient algorithm was applied, and the energy cutoff was set at 500 eV. Reciprocal space integrations were completed over a 8×8×2 Monkhorst-Pack *k*-point mesh.[36] SOC effects were applied for both Ru and I atoms. Orbital potentials (U = 4 eV) were employed for Ru $d$ electrons based on previous calculations for $\alpha$-RuCl$_3$.[29] The structural parameters for RuI$_3$ were obtained from experiment, and the parameters for RuCl$_3$ were obtained from the ICSD.[21] With these settings, the calculated total energy converged to less than 0.1 meV per atom.

[CSD 2105773-2105774 contains the supplementary crystallographic data for this paper. These data can be obtained free of charge from The Cambridge Crystallographic Data Centre via www.ccdc.cam.ac.uk/data_request/cif.]

**Supporting Information**

Supporting Information is available from the Wiley Online Library or from the author.

**Acknowledgement**

The research reported here was funded by the Gordon and Betty Moore foundation, EPiQS initiative, grant GBMF-9006. K.M.P. acknowledges the support of the NSF Graduate Research Fellowship under Grant No. DGE-1656466. The authors thank Prof. Weiwei Xie at Rutgers University for the use of her single crystal X-ray diffractometer. The authors acknowledge the use of Princeton's Imaging and Analysis Center, which is partially supported by the Princeton Center for Complex Materials, a NSF-MRSEC program (DMR-1420541).

**Conflict of Interest**

The authors declare no conflict of interest.

**Table 1.** Atomic coordinates and equivalent isotropic displacement parameters for $Ru_{0.97(1)}I_3$ at 300 (2) K. ($U_{eq}$ is defined as one-third of the trace of the orthogonalized $U_{ij}$ tensor (Å$^2$))

| Atom | Wyck. | Occ. | x | y | z | $U_{eq}$ |
|---|---|---|---|---|---|---|
| I1 | 18f | 1 | 0.3510 (1) | 0.0014 (1) | 0.07960 (3) | 0.0175 (2) |
| Ru2 | 6c | 0.957 (7) | 0.6667 | 0.3333 | 0.9998 (1) | 0.0144 (4) |
| Ru3 | 3a | 0.021 (6) | 0 | 0 | 0 | 0.014 (2) |

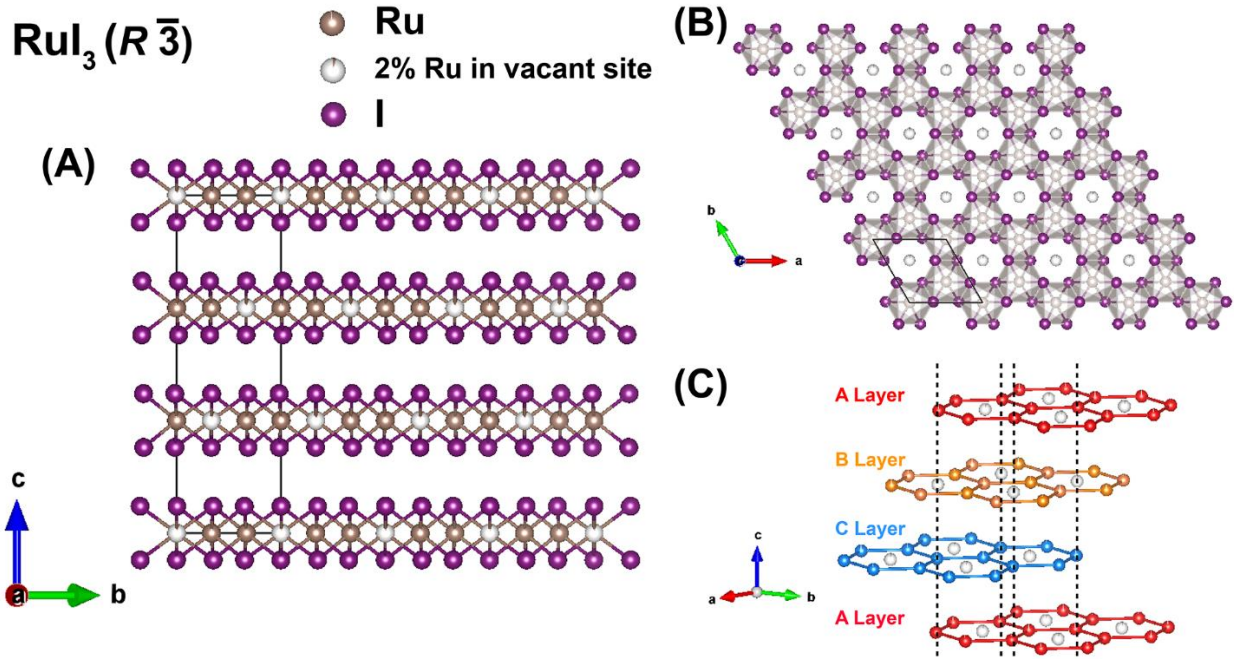

**Figure 1.** Crystal structure of honeycomb-structure α-RuI$_3$, viewed along (A) the *a*-axis and (B) the *c*-axis. The RuI$_6$ octahedra form a layered honeycomb lattice by sharing edges, and the honeycomb layers are stacked in an A B C sequence as shown in (C). The white circles in A and B (occupied at a 2% Ru fraction) show the location of the partial occupancy of Ru on the normally empty sites in the honeycomb lattice, tentatively ascribed to the presence of a very small number of stacking faults.

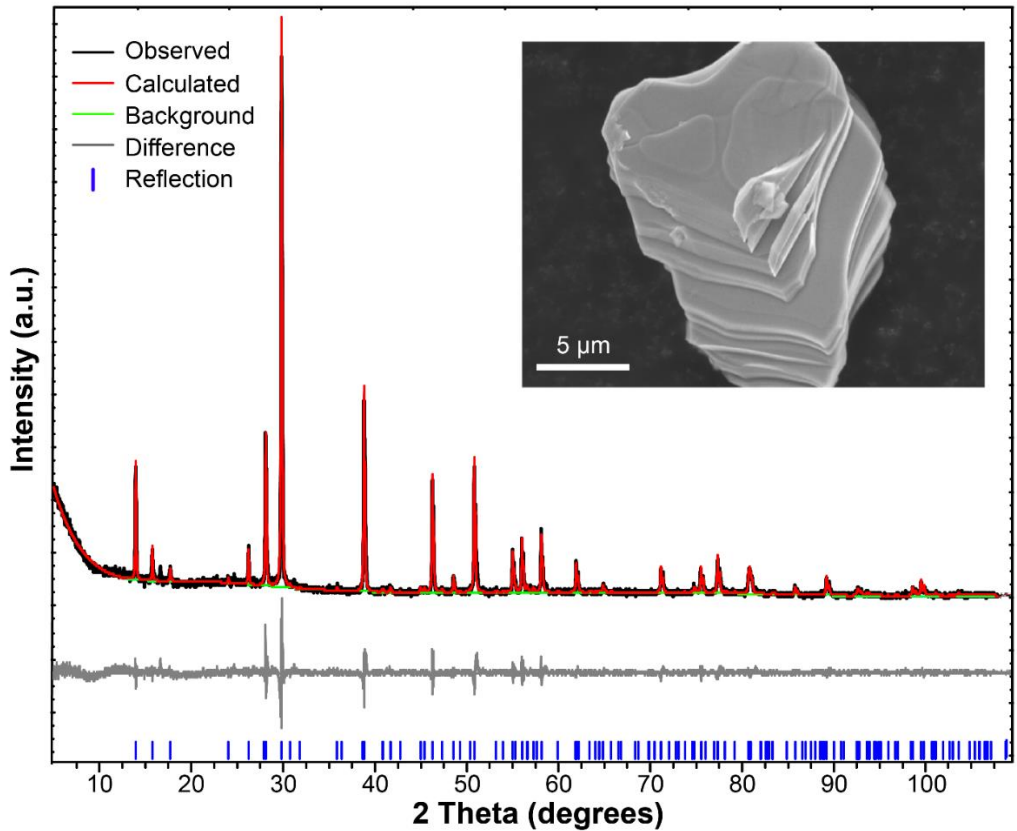

**Figure 2.** PXRD pattern showing a Le Bail fit for $\alpha$-RuI$_3$, confirming the consistency of the bulk sample with the SXRD result. An SEM image of a grain of the material is shown in the inset, supporting its layered nature.

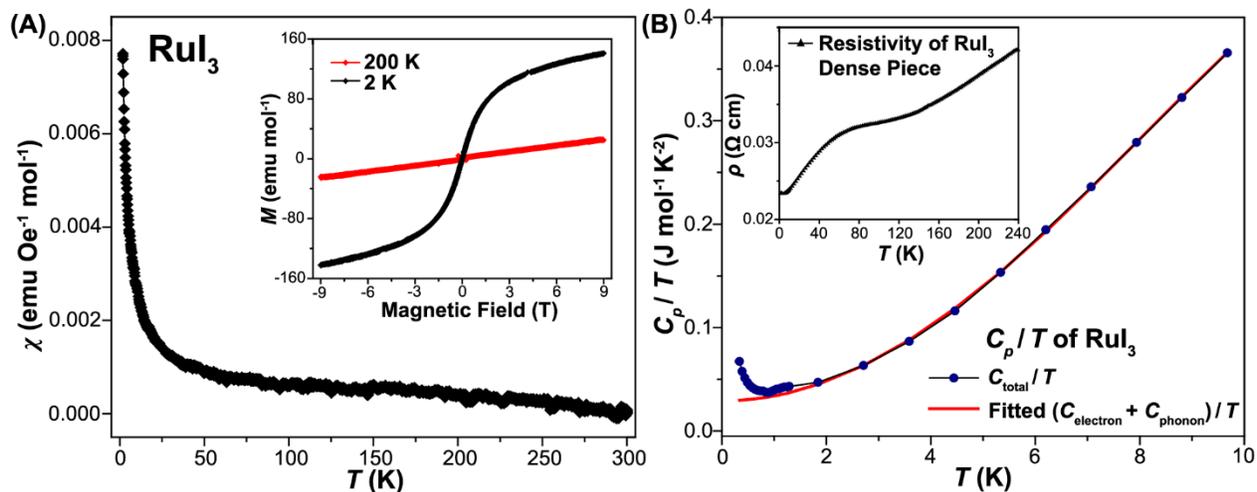

**Figure 3.** General magnetic and electronic characterization of $\alpha$-RuI$_3$. (A) Magnetic susceptibility measured from 1.8 to 300 K. The inset presents the magnetization measured versus applied magnetic field (-9 T to 9 T) at 2 K and 200 K, respectively; (B) heat capacity measured from 0.35 K to 10 K. The inset shows the preliminary temperature-dependent resistivity measurement on an as-made dense piece of $\alpha$-RuI$_3$.

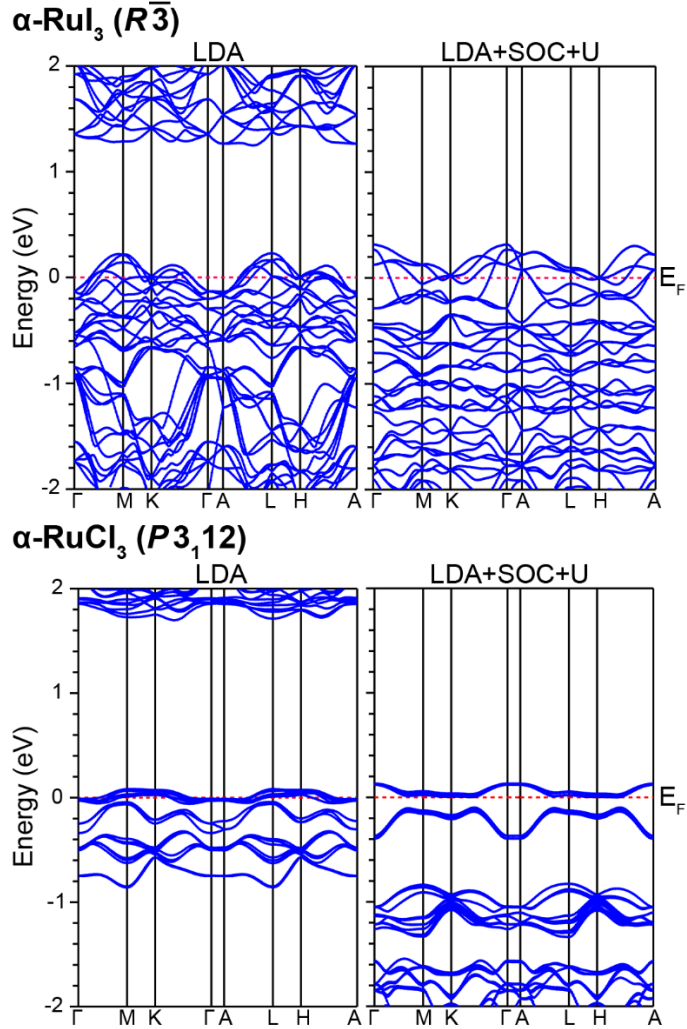

**Figure 4.** The calculated electronic band structures of α-RuI₃ (A) without and (B) with electron repulsion and spin orbit coupling (SOC) included. Same story for α-RuCl₃ (C) and (D). For α-RuCl₃ the information for doing the calculations was taken from the $P3_112$ structure CIF file in the International Crystal Structure Database (ICSD).[21] The U parameter for RuCl₃,[29] was employed in both cases.

16